\documentclass[preprint,showpacs,preprintnumbers,amsmath,amssymb]{revtex4}
\usepackage{graphicx}
\usepackage{dcolumn}
\usepackage{bm}
\usepackage{amsfonts}
\usepackage{amsmath}
\usepackage{textcomp}
\usepackage{amsxtra}
\usepackage{mathtools, bm}
\usepackage{amssymb, bm}
\usepackage{color}

\begin{document}

\title{Twofold stationary states in the classical Spin-Hall effect}
\author{J.-E. Wegrowe} \email{jean-eric.wegrowe@polytechnique.edu}
\affiliation{Ecole Polytechnique, LSI, CNRS and CEA/DSM/IRAMIS, Palaiseau F-91128, France} 
\date{\today}

\begin{abstract}
The stationary states occurring in spin-Hall devices are investigated within the framework of the phenomenological two spin-channel model. It is shown that two different stationary states can be defined, that depends on the redistribution of the electric charges between the two spin-channels during the transient time. A first stationary state can be reached if the charge accumulation occurs inside each spin channel independently, while a second stationary state is reached if the two spin channels are undifferentiated from the point of view of the electric charge accumulation. The screening equations that describe the accumulation of electric charges due to spin-orbit coupling are derived in both cases, and the two stationary states are discussed in terms of the Dyakonov-Perel transport equations. It is shown that a fictitious spin-dependent electric field should be introduced in the equations in order to take into account the first stationary state. In both cases, the phenomenology is compatible with experimental observations. 
\end{abstract}
\maketitle 

\vspace{2.5cm}

\section{Introduction}

The spin-Hall effect (SHE) that occurs in non-ferromagnetic materials with strong spin-orbit interaction is one of the most important effect in the new field of spin-orbitronics. However, in spite of a large number of excellent reports about SHE (see \cite{Dyakonov,Dyakonov2,Hirsch,Zhang,Tse,Maekawa,Hoffmann,Review,Saslow} and references therein), a fundamental ambiguity seems to persist about the definition of the stationary regime, leading to different predictions as to the presence of a transverse pure spin-current \cite{EPL2017}. 

The SHE describes the spin-accumulation generated at the transverse edges of a non-ferromagnetic conductor with strong spin-orbit coupling (SOC), while injecting a longitudinal electric current \cite{Awschalom,Wunderlich,Valenzuela,Gambardella,Otani}. The phenomenological description of the SHE is based on the Dyakonov-Perel transport equations \cite{Dyakonov,Dyakonov2}, which are a generalization of Ohm's law for the spin-dependent electric carriers in the presence of SOC. In a thin conducting layer, the SOC generates a spin-polarization of the electric carriers, that can be described by two channels, one for the spin up $\uparrow$ and the other for the spin down $\downarrow$ with a spin-polarization oriented along the axis perpendicular to the plane of the layer. The SOC plays then the role of an effective magnetic field (see Fig.1), that acts independently on the two populations of electric charges. The system can then be viewed as a simple superimposition of two standard Hall devices composed of two populations of electric charges experiencing opposite magnetic fields $\vec H_{so \updownarrow} = \pm \vec H_{so}$. According to the usual Hall effect, the magnetic field along the $\vec H_{so}$ direction generates a charge accumulation $\delta n_{\uparrow}$ at the edges, while the magnetic field along the $-\vec H_{so}$ direction generates a charge accumulation of opposite sign $\delta n_{\downarrow} = - \delta n_{\uparrow}$. 

In a first approach, it can be conclude that the superimposition of the two sub-systems leads to a vanishing total electric charge accumulation at the edges $\delta n = \delta n_{\uparrow} + \delta n_{\downarrow} = 0$ (but with non-zero spin-accumulation $\Delta n = \delta n_{\uparrow} - \delta n_{\downarrow} = 2 \delta n_{\uparrow}$). A stationary pure spin-current $J_{y \uparrow} = - J_{y \downarrow} \ne 0$ is then generated in the transverse direction \cite{Zhang,Tse,Maekawa,Hoffmann,Review,Saslow,EPL2017}. However, if the time necessary to establish the two spin-channels - i.e. the spin-orbit scattering time $\tau_{so}$ - is much shorter than the spin-flip relaxation time, the stationary state can be reached inside each spin channels independently. The charge accumulation in each spin channel hence produces a transverse electric field $E_{y \updownarrow}$ in each spin channel, such that $E_{y \uparrow} = - E_{y \downarrow}$ (this spin-dependent electric field is depicted  in the paper of Hirsch \cite{Hirsch} in the right bottom of Fig.1). In that case, like in the usual Hall effect, the transverse stationary current $J_{y \updownarrow} = 0$ is zero for both spin channels. 

There is hence {\it two different options about the stationary states}. The first stationary state with spin-dependent electric currents dissipates more than the second one with spin-dependent electric fields. 
Since the second law of thermodynamics imposes that the stationary state corresponds to the minimum heat dissipation under the constraints imposed to the system, the first system is more constrained than the second one. As shown in a recent work based on this variational principle \cite{EPL2017} the minimum power dissipated in the spin-Hall system is indeed reached for $J_{y \uparrow} = J_{y \downarrow} = 0$ in the bulk (without the need to assume a spin-dependent electric field), except for Corbino configurations.

  \begin{figure} [h!]
   \begin{center} 
   \begin{tabular}{c}
   \includegraphics[height=10cm]{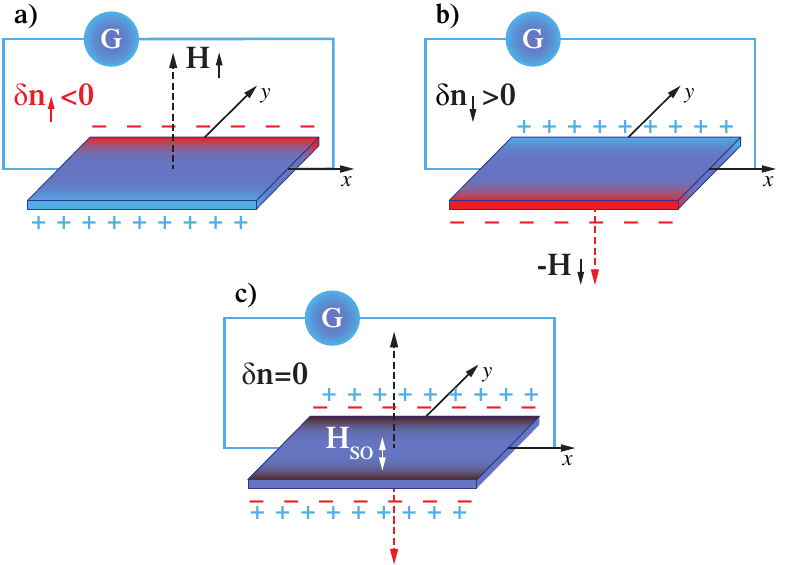}
   \end{tabular}
   \end{center}
   \caption[Fig2]
{ \label{fig:Fig2} : Schematic representation of the spin-Hall effect with the electrostatic charge accumulation $\delta n_{\updownarrow}$ at the boundaries: (a) usual Hall effect with a magnetic field $\vec H_{\uparrow}$ in the up direction (b) usual Hall effect with a magnetic field $\vec H_{\downarrow}$ in the down direction, (c ) the addition of  configuration (a) and (b) leads to the description of the spin-orbit scattering with an effective magnetic field $\vec H_{so}$ acting on the two different electric carriers}
   \end{figure}

The aim of this work is to study these twofold stationary states of the SHE, and to understand the consequences in terms of screening equations, spin-accumulation equation and transport equations. The approach is based on the application of mesoscopic non-equilibrium thermodynamics  \cite{Rubi} to spin-dependent transport \cite{JPhys07}, in which local equilibrium is assumed. The paper is structured as follow: the section II treats the standard Hall effect in the formalism of mesoscopic non-equilibrium thermodynamics. The screening equation is derived and the usual approximations related to constant conductivity and small Debye-Fermi length are analyzed. The section III is devoted to the two spin-channel model for the SHE, and the derivation of the screening equations  in the absence spin-flip scattering. The section IV presents the derivation of the screening equation and spin-accumulation in the case of spin-flip scattering, and section V analyzes the corresponding transport equations in terms of the Dyakonov-Perel equations.

\section{Hall effect}

The goal of this section is to characterize the stationary state of the standard Hall device on the basis of the stationarity condition $dn/dt = 0$ (where $n$ is the density of electric carriers), and investigating the approximations that consists in assuming constant conductivity and small screening length. 

Ohm's law reads:
\begin{equation}
\vec J = - \hat \eta n \vec  \nabla \mu,
\label{OhmN}
\end{equation}
where $n$ is the density of electric carriers of charge $q$, $\mu$ is the electro-chemical potential, and the mobility tensor $\hat \eta$ is related to the conductivity tensor $\hat \sigma$ by the relation $\hat \eta = \hat \sigma/(q n)$. If the sample under consideration is an isotropic planar layer, on which a magnetic field $\vec H$ is applied along the direction $\vec e_z$ perpendicular to the layer, the mobility tensor in the orthonormal basis $\{ \vec e_x, \vec e_y, \vec e_z \}$ is expressed by the matrix:
\begin{equation}
\hat \eta =
\left( \begin{array}{cc}
                       \eta & \eta_{H} \\
                      - \eta_{H}  &  \eta \\
                    		\end{array} \right),               
  \label{Mobility}
\end{equation}
where the coefficient $\eta_H$ is the Hall mobility (which is an odd function of $H$). According to the Drude relation we have $\sigma = q^2 n \tau / m^*$ (where $\tau$ is the electronic relaxation time and $m^*$ is the effective mass) so that $\eta = q \tau/m^*$ does not depend on $n$.

Introducing a unit vector $\vec p$ in the direction of the applied field $\vec H$, the vector form of Eq.(\ref{OhmN}) reads:
\begin{equation}
\vec J = - \eta n  \vec \nabla \mu  + \vec p \times n \eta_{H} \vec \nabla \mu.
\label{OhmN2}
\end{equation}
On the other hand, the chemical potential is defined by :
\begin{equation}
\mu= kT ln(n/n_0) + V_G + V(\vec H) + \mu_{0},
\label{ChemPot}
\end{equation}
where $T$ is either the Fermi temperature ($T = T_F$) in the case of a metal or the temperature of the thermostat in the case of a non-degenerated semi-conductor, $k$ is the Boltzmann constant, $n_0$ is the constant density of the electric carriers that corresponds to the electric neutrality of the material, and $\mu_{0}$ describes the chemical potential related to internal degrees of freedom. The electric potential $V_G$ is produced by the electric generator and imposes the constant electric field $qE_x^0 = -\partial V_G/\partial x$ along the $x$ axis. On the other hand, the electric potential $V(\vec H)$ takes into account the effect of the magnetic field $\vec H$. Due to the Lorentz force, the consequence of the application of the magnetic field is a redistribution of the electric carriers in the direction $y$, which results in a charge accumulation  $\delta n = n - n_0$ at the edges. The charge accumulation generates, in turn, an electric field $q E_y = -  \partial V(\vec H)/\partial y$ defined by Gauss's law:
 \begin{equation}
\frac{ \partial E_y}{\partial y} = \frac{q\delta n} \epsilon
 \label{PoissonEq}
 \end{equation}
For convenience, the two contributions $\vec \nabla V_G$ and $\vec \nabla V(H)$ of the gradient of the chemical potential can be combined in a single electric field $q \vec E =- \vec \nabla V_G - \vec \nabla V(\vec H)= q E_x^0 \vec e_x + q E_y \vec e_y$. Inserting the expression of the chemical potential Eq.(\ref{ChemPot}) into Ohm's law Eq.(\ref{OhmN2}) we obtain:
\begin{equation}
\vec J =   qn \eta \vec E - D \vec \nabla n -  \vec p \times \left (qn \eta_H \vec E - D_H \vec \nabla n  \right),
\label{OhmN3}
\end{equation}
where the diffusion constants are defined as $D= \eta kT$ and $D_H= \eta_H kT$. 

The divergence of Eq.(\ref{OhmN3}) reads:
\begin{equation}
div (\vec J) =  q n \eta \, div(\vec E ) + q \eta \vec \nabla n . \vec E - D \nabla^2 n,
\label{divOhmN}
\end{equation}
where we have used the relations $div(n \vec E) = \vec E . \vec \nabla n + n \, div(\vec E)$ and $div(\vec p \times \vec \nabla \mu) = \vec \nabla \mu . \vec{rot}( \vec p) - \vec p . \vec{rot}( \vec \nabla \mu) = 0$. The consequence of the last relation is that the Hall term in the right hand side of Eq.(\ref{OhmN3}) disappears from the definition of the stationary states. Nevertheless, the Hall effect is still present as it is responsible for the charge accumulation $\delta n$ at the edges.
The stationarity condition writes $\partial n/ \partial t = -div \vec( J) = 0$. In the case of a recombination between electrons and holes, or for other relaxation mechanisms, we should add the relaxation parameter $\dot R$ such that $div \vec (J) = -\dot R$ (an explicit expression of $\dot R$ can be found in \cite{Tse}). The stationarity condition leads to the following {\it screening equation} for the electric charge accumulation $\delta n$: 
\begin{equation}
\frac{\partial^2 \delta n}{\partial y^2}  - \frac{\delta n}{\lambda_D^2} = \frac{\partial \delta n}{\partial y}  \frac{q^2}{\epsilon kT} \int_y \delta n (y') dy' + \frac{\dot R}{D}
\label{Screening}
\end{equation}
where $\lambda_D = \sqrt{\frac{\epsilon kT}{q^2 n}}$ is the screening length. This is an exact result for the Hall bar assuming translation invariance along $x$. In the case $n \gg \vert \delta n \vert$, the screening length is the Debye length $\lambda_D \approx \bar \lambda_D \equiv  \sqrt{\frac{\epsilon kT}{q^2 n_0}}$.

\subsection {Constant conductivity approximation}

When $\vert \delta n \vert \ll n_0$, the conductivity $\sigma = q \eta (n_0 + \delta n)  \approx q \eta n_0$ is constant. If we further assume that $\dot R \approx 0$, then Eq.(\ref{Screening}) reduces to the well-known screening equation at equilibrium:
\begin{equation}
\nabla^2 \delta n - \frac{\delta n}{\lambda_D^2} \approx 0.
\label{DiffEq}
\end{equation}
The electrostatic charge accumulation decays exponentially to zero with a typical length $\bar \lambda_D$. Consequently, the last term $q \eta \vec \nabla n . \vec E$ in the left hand side of Eq.(\ref{divOhmN}) accounts for the variation of the conductivity $\sigma$ due to the electrostatic charge accumulation over a distance $\lambda_D$ from the border of the Hall device. The approximation $\vert \delta n \vert \ll n_0$ as been well established as correct for conventional systems.

\subsection{Small Debye-Fermi length and bulk approximation}

If we assume $\lambda_D$ small, we have in second order:
\begin{equation}
- \delta n = \frac{\partial \delta n}{\partial y}  \left ( \frac{1}{n } \int_y \delta n (y') dy' \right ) \approx \frac{\partial \delta n}{\partial y}  \left ( \frac{1}{n_0 } \int_y \delta n (y') dy' \right ),
\label{Screening2}
\end{equation}
A moderate non-conservation of the electric charges $\dot R \ne 0$ has no consequence in the framework of this approximation.
\subsection{Bulk approximation for the stationary state}
The approximation of small Debye-Fermi length  {\bf B} together with that of constant conductivity {\bf A} (the right-hand side of Eq.(\ref{Screening2}) is vanishing) results in the {\it bulk approximation} of the stationary state: the charge accumulation $\delta n(y)$ reduces to surface charges at both edges (for the Hall bar geometry). In our $2D$ model, we have two opposite Dirac distributions located at $y = \pm l$ where $l$ is the width of the Hall device. The charge accumulation is zero inside the device $\forall y \gg \lambda_D,$ $\delta n(y) = 0 $, the transverse stationary current is zero $J_y = 0$, the electric field produced by the surface charges $E_y$ is constant and the Hall potential $V(y) \propto y$ is linear. 

\section{Spin-Hall effect in the two spin channel model}

The goal of this section is to introduce the two spin channels labeled by the index $\updownarrow$, and to characterize the stationary states defined by the conservation laws  $dn_{\updownarrow}/dt = -div \vec J_{\updownarrow} - \dot R = 0$ in the absence of spin-flip scattering. The two channels are generated by the SOC. It is assumed that the spin-orbit scattering time is much shorter than the spin-flip relaxation time. Like for the description of the giant magnetoresistance in the two spin-channel model \cite{Johnson,Wyder,Valet-Fert,PRB2000,JPhys07,Tulapurkar}, the two sub-systems can then be treated separately, and the spin-flip relaxation is introduced through the conservation laws for the electric carriers in a second step (see next section).

The spin-dependent conductivity $\sigma_{\updownarrow}$ of the spin-channels can be introduced through the Drude formula: $\sigma_{\updownarrow} = q^2 n_{\updownarrow} \tau/m^*$ where the relaxation $\tau$ is not spin-dependent, and $m^*$ is the effective mass. Hence, the mobility $\eta = \sigma_{\updownarrow}/(q n_{\updownarrow})$ does not depend on $n_{\updownarrow}$ and is spin independent. Consequently, the diffusion constant is also spin-independent $D= \sigma_{\updownarrow} kT/n_{\updownarrow} = \eta kT$. In contrast, the non-diagonal mobility coefficients $\eta_{so \updownarrow} = \pm \eta_{so}$ due to spin-orbit scattering and the corresponding diffusion constant $D_{so \updownarrow} = \eta_{so \updownarrow} kT = \pm \eta_{so} kT $ are spin-dependent (the sign ($+$) corresponds to the spin $\uparrow$ and the sign ($-$) corresponds to the spin $\downarrow$). Ohm's law for each spin-channel reads:
\begin{equation}
\begin{array}{c}
\vec J_{\uparrow}= - \hat  \eta_{\uparrow} n_{\uparrow} \vec  \nabla \mu_{\uparrow}\\
\vec J_{\downarrow}= - \hat  \eta_{\downarrow} n_{\downarrow} \vec  \nabla \mu_{\downarrow}.
\end{array}
\label{OhmS}
\end{equation}
Let us write more concisely the two equations in the form $\vec J_{\updownarrow}= - \hat  \eta_{\updownarrow} n_{\updownarrow} \vec  \nabla \mu_{\updownarrow}$, where the mobility tensor $ \eta_{\updownarrow}$ is given by the matrix:
\begin{equation}
\hat \eta_{\updownarrow} =
\left( \begin{array}{cccc}
                       \eta & \eta_{so} & 0 & 0  \\
                      - \eta_{so}  &  \eta & 0 & 0 \\
                      0 & 0 &  \eta & -\eta_{so} \\
                      0 & 0 & \eta_{so}  &  \eta \\
		\end{array} \right).                
  \label{MobilityS}
\end{equation}
In a vector form we have the two equations:
\begin{equation}
\vec J_{\updownarrow}= - \eta n_{\updownarrow}  \vec \nabla \mu_{\updownarrow}  \pm \vec p \times n_{\updownarrow} \eta_{so} \vec \nabla \mu_{\updownarrow}
\label{OhmS2}
\end{equation}
where the sign ($+$) corresponds to the spin $\uparrow$ and the sign ($-$) corresponds to the spin $\downarrow$. 
As shown in section V below, Eq.(\ref{OhmS2}) is a generalization of the Dyakonov-Perel equations. 

According to the model depicted in Fig.1, the chemical potential of the charge carriers in each spin channel is given by Eq.(\ref{ChemPot}):
\begin{equation}
\mu_{\updownarrow} = kT ln(n_{\updownarrow}/n_0 ) + V_G + V(H_{so \updownarrow}) + \mu_{0 \updownarrow},
\label{ChemPotS}
\end{equation}
where $V_{G}$ is the contribution of the electric generator which imposes a constant electric field along the $x$ direction: $q E_x^0 = - \partial V_G/\partial x$. The contribution $V(H_{so \updownarrow})$ takes into account the effect of SOC through the effective magnetic field $H_{so \updownarrow}$, and $\mu_{0 \updownarrow}$ accounts for the spin-dependent properties of the electric carriers. 

As described in the tensor Eq.(\ref{MobilityS}), the effect of the effective magnetic field $H_{so \updownarrow} = \pm H_{so}$ is equivalent to that of the usual Hall effect for each spin-channel with a internal magnetic field applied in opposite direction. It leads to a redistribution of the electric charges $\delta n_{\updownarrow}$ such that $\delta n_{\uparrow} = - \delta n_{\downarrow}$. This distribution of electric charges defines - in each spin channel - a spin-dependent electric field $E_{y \updownarrow}$ along the $y$-direction through Gauss's law $\partial E_{y \updownarrow}/ \partial y = q \delta n_{\updownarrow}/\epsilon$ (in other terms, the introduction of a spin-dependent charge accumulation $\delta n_{\updownarrow}$ is equivalent to a spin-dependent electric potential $\partial E_{y \updownarrow}/ \partial y$). The two contributions $V_G$ and $V(H_{so \updownarrow})$ in Eq.(\ref{ChemPotS}) can be combined into a single electric field $ q \vec E_{\updownarrow} = - \vec \nabla V_{\updownarrow} =  - \vec \nabla V_G - \vec \nabla V(H_{so \updownarrow}) = qE_x^0 \vec e_x + q E_{y \updownarrow} \vec e_y$. Inserting the expression of the chemical potential Eq.(\ref{ChemPotS}) into the Omh's law Eq.(\ref{OhmS2}) yields:
\begin{equation}
\vec J_{\updownarrow} =   qn_{\updownarrow} \eta \vec E_{\updownarrow} - D \vec \nabla n_{\updownarrow} \mp  \vec p \times \left (qn_{\updownarrow} \eta_{so} \vec E_{\updownarrow} - D_{so} \vec \nabla n_{\updownarrow}  \right),
\label{GradChemPotS}
\end{equation}
where $D_{so} = kT \eta_{so}$, and we assume that $\mu_{0 \updownarrow}$ is constant due to the absence of spin-flip scattering (this assumption is removed in the next section). It is worth pointing-out that the a naive superimposition of the two sub-systems (a) and (b) of Fig.1 should leads to a superimposition of the equal and opposite magnetic fields, i.e. should lead to a vanishing total magnetic field for the total system (c). This is of course not the case, because the two magnetic fields are ``local" in the spin space, in the sense that the two values are taken at different ``positions'' in the spin configuration space (which is reduced to the north and south poles of the Bloch sphere for the two spin channel model). 

On the other hand, if the two sub-systems of Fig.1 are mixed before the stationarity condition is reached, we have $\delta n = 0$, and the local electric fields $\vec E_{\updownarrow}$ cannot be defined. The corresponding system is then analogous to the Corbino configuration \cite{EPL2017} in which charge accumulation is forbidden, and this configuration imposes a Hall current for both subsystems in the stationary state: we have $J_{y \uparrow} = - J_{y \downarrow}$, instead of $J_{y \uparrow} = J_{y \downarrow} = 0$. In agreement with the Dyakonov-Perel equations (see section V below), most publications about the SHE \cite{Review} assume the more constrained stationary state $J_{y \uparrow} = - J_{y \downarrow}$ with the generation of a pure spin-current. We focus here on the less constrained system $\delta n_{\uparrow} = -\delta n_{\downarrow}$, keeping in mind that the previous stationary state is recovered with putting $E_y = E_{y \uparrow} + E_{y \downarrow}  = 0$, i.e. having $\vec E$ instead of $\vec E_{\updownarrow}$. 

From Eq.(\ref{OhmS2}), we derived the expression of the stationarity condition $div (\vec J_{\updownarrow}) = - \dot R/2$ :
\begin{equation}
- D \nabla^2 n_{\updownarrow}  + q \eta n_{\updownarrow} \, div \vec E_{\updownarrow}  + q \eta \vec \nabla n_{\updownarrow} . \vec E_{\updownarrow} = -\dot R/2.
\label{divOhmS2}
\end{equation}
Equation (\ref{divOhmS2}) can be put into the form:
\begin{equation}
\frac{\partial^2  \delta n_{\updownarrow}}{\partial y^2} - \frac{\delta n_{\updownarrow}}{\lambda_{D \updownarrow}^2} - \frac{q}{kT} \vec \nabla \delta n_{\updownarrow} . \vec E_{\updownarrow} = \frac{\dot R}{2D},
\label{divOhmS3}
\end{equation}
where the screening length $\lambda_{D \updownarrow} = \sqrt{\frac{\epsilon kT}{q^2 n_{\updownarrow}} }$ is approximatively equal to the Debye length $\lambda_D \approx \sqrt{\frac{\epsilon kT}{q^2 n_0} }$. 

Using $\partial E_{\updownarrow}/ \partial y = q \delta n_{\updownarrow}/\epsilon$, equation (\ref{divOhmS3}) reduces to :
\begin{equation}
 \lambda_{D  \updownarrow}^2 \, \left ( \frac{\partial^2  \delta n_{\updownarrow} }{\partial y^2} - \frac{\dot R}{2D} \right ) = \delta n_{\updownarrow} + \frac{\partial  \delta n_{\updownarrow}}{\partial y}  \frac{1}{n_{\updownarrow}} \, \int_y    \delta n_{\updownarrow} \, dy' 
\label{ScreeningS}
\end{equation}
For each spin channel, Eq.(\ref{ScreeningS}) is equivalent to that of the simple Hall effect Eq.(\ref{Screening}) because the two spin-dependent equations are decoupled. This is also equivalent to a simple screening equation of a transport process without Hall effect, in which the charge accumulation $\delta n_{\updownarrow}$ is imposed at the interfaces by other means. Note however that for the other stationary state (for which the spin-dependent electric fields $E_{y \updownarrow}$ do not exist), term $\delta n_{\updownarrow}$ in the right-hand side should be replaced by $\delta n$ (the first term and the term inside the integral), and the two equations are then coupled.

In the approximation in which we assumed {\it constant conductivity} ($\vert \delta n \vert \ll n$) and $\dot R \approx 0$  we obtain:
\begin{equation}
\frac{\partial^2  \delta n_{\updownarrow} }{\partial y^2}  - \frac{\delta n_{\updownarrow} }{\lambda_D^2} \approx 0,
\label{DiffEq}
\end{equation}
The effect of the electrostatic charge accumulation decreases exponentially over the screening length $\lambda_D$ :$\delta n_{\updownarrow} \sim \delta n_{0 \updownarrow} \, e^{- y/\lambda_D}$ where $\delta n_{0 \updownarrow}$ is the charge accumulation at the edges produced by the effective field $\vec H_{so \updownarrow}$. The spin accumulation takes place $\Delta n = \delta n_{\uparrow} - \delta n_{\downarrow} = 2 \delta n_{\uparrow} \ne 0$ with zero total charge accumulation $\delta n = 0$ \cite{Dyakonov2,Review}. 

On the other hand, in the approximation of small Debye length but without assuming constant conductivity, entails the relation:
\begin{equation}
 - \delta n_{\updownarrow} = \frac{\partial  \delta n_{\updownarrow}}{\partial y}  \frac{1}{n_{\updownarrow}} \, \int_y    \delta n_{\updownarrow} \, dy' .
\label{ScreeningS4}
\end{equation}
The approximation of small Debye-Fermi length together with that of constant conductivity (the right-hand side of Eq.(\ref{Screening2}) is vanishing) results in the {\it bulk approximation} $\delta n_{\updownarrow} \approx 0$ and $\Delta n \approx 0$. In this usual approximation, this result is equivalent to the other stationary state with spin-independent field $\vec E_{\updownarrow} = \vec E$ (or $V(H_{so \updownarrow}) = V$ in Eq.(\ref{ChemPotS})). Consequently, from the point of view of the charge accumulation, {\it the two different stationary states discussed in the introduction ($J_{y \updownarrow} = 0$ vs. $E_{y \updownarrow} = 0$) can no longer be discriminated in the bulk approximation}.

\section {Spin-Hall effect with spin-flip relaxation}

In the description used so far, we assumed that the two sub-systems that are generated by the spin-orbit scattering can be described independently. This means that the spin-flip relaxation time $\tau_{sf}$ is much larger than both the spin-orbit scattering time $\tau_{so}$ and the transient time, but it does not mean that the spin-flip relaxation time is infinite.
  
It is easy to take into account phenomenologically the spin-flip relaxation in the framework of the two channel model, on the basis of relaxation of the internal degrees of freedom \cite{JPhys07,PRB2000,Entropy,Tulapurkar}. The two spin-populations are put out-of equilibrium by the SOC ($\Delta \mu \ne 0$), and the spin-flip relaxation $\uparrow \, \overset{\dot \psi}{\longrightarrow} \, \downarrow$ can be treated as a chemical reaction that transforms a conduction electron of spin $\uparrow$ into a conduction electron of spin $\downarrow$ at the rate $\dot \psi$. The power dissipated by the spin-flip relaxation process is $P_{sf} = \dot \psi \Delta \mu$, where the chemical affinity $\Delta \mu$ corresponding to this reaction is $\Delta \mu = \mu_{\uparrow} - \mu_{\downarrow}= kT ln(\frac{n_{\uparrow}}{n_{\downarrow}}) - q \frac{\partial (E_{y \uparrow} - E_{y \downarrow})}{\partial y} + \Delta \mu_0$, where $\Delta \mu_0 = \mu_{0 \uparrow} - \mu_{0 \downarrow}$. The reaction rate $\dot \psi$ is defined by the conservation laws as $dn_{\updownarrow}/dt = - div J_{\updownarrow} \mp \dot \psi$. A supplementary transport equation in the spin configuration-space describes the spin-flip relaxation process \cite{Johnson,Wyder,Valet-Fert,PRB2000,JPhys07,Tulapurkar} 
\begin{equation}
\dot \psi = \mathcal L \, \Delta \mu,
\label{SpinFlip}
\end{equation}
 that relates the flux $\dot \psi$ to the force $\Delta \mu$, where the transport coefficient $\mathcal L \propto 1/\tau_{sf}$ is positive and inversely proportional to the spin-flip relaxation time \cite{JPhys07,Tulapurkar}. 

Furthermore, due to spin-flip scattering, the spin-dependent chemical potential $\mu_{0 \updownarrow}$ is no longer constant, and we have $\partial \mu_{0 \updownarrow}/\partial y \ne 0$. In the framework of the two channel model, the stationarity condition becomes:
\begin{equation}
div (\vec J_{\updownarrow}) = -\frac{\dot R}{2} \mp \mathcal L \Delta \mu
\label{SpinCons}
\end{equation}
 and the screening equation Eq.(\ref{divOhmS3}) now reads:
\begin{equation}
  \frac{\partial^2 \delta n_{\updownarrow}}{\partial y^2} - \frac{\delta n_{\updownarrow}}{\lambda_{D \updownarrow}^2} -  \frac{\partial \delta n_{\updownarrow}}{\partial y}  \frac{1}{\lambda_{D \updownarrow}^2 n_{\updownarrow}}   \int_y \delta n_{\updownarrow} \, dy' + \frac{1}{kT} \left (n_{\updownarrow} \frac{\partial^2  \mu_{0 \updownarrow}}{\partial y^2} + \frac{\partial  \mu_{0 \updownarrow}}{\partial y} \, \frac{\partial  \delta n_{\updownarrow}}{\partial y}   \right ) =  \frac{1}{2 \eta kT} \left (\dot R  \pm 2 \mathcal L \Delta \mu \right ).
\label{Screeningsf2}
\end{equation}
We will not try to solve this system of two coupled equations, but the analysis performed in the previous sections remains applicable. 

Let us focus on the {\it bulk approximation} for which $\delta n_{\updownarrow}$ is reduced to a surface charge at both edges (i.e. we have two Dirac peaks at both edges $\delta n_{\updownarrow}(y) \propto \pm (\delta(y + l) + \delta(y - l))$). This approximation is also valid in the region $\lambda_D \ll y \ll l_{sf}$.  Assuming, for the sake of simplicity, that $\dot R \approx 0$, Eq. (\ref{Screeningsf2}) then reduces to: 
\begin{equation}
 \frac{\partial^2  \mu_{0 \updownarrow}}{\partial y^2}  = \pm \frac{\Delta \mu}{l_{sf}^2}.
\label{SpinDiff}
\end{equation}
where we have defined the spin diffusion length in our non-ferromagnetic conductor by $l_{sf} =  \sqrt{\sigma_0/(2q \mathcal L)}$. Since, in this approximation the chemical potential is linear in the absence of spin-flip scattering, we obtain the spin-accumulation equation for $\Delta \mu$, as defined in the context of the GMR \cite{Johnson,Wyder,Valet-Fert,PRB2000,JPhys07,Tulapurkar,Shibata}:
\begin{equation}
 \frac{\partial^2 \Delta \mu}{\partial y^2} = 2 \frac{\Delta \mu}{l_{sf}^2},
 \label{GMR}
 \end{equation}
 This analysis shows that, providing that the condition $l_{sf} \gg \lambda_D$ is satisfied, the spin-accumulation $\Delta \mu(y)$ due to the SHE decreases over the spin-flip diffusion length $l_{sf}$ in the same way than GMR spin-accumulation, as observed experimentally \cite{Awschalom,Wunderlich,Valenzuela,Gambardella}.
 
Note that the result Eq.(\ref{GMR}) could be { \it conter-intuitive} at first glance because the spin-accumulation $\Delta n = \delta n_{\uparrow} - \delta n_{\downarrow}$ is produced by the charge accumulation only, which characteristic length is $\lambda_D$ (i.e. about zero) and not $l_{sf}$. The reason for this apparent paradox is the same as for the usual Hall effect, for which the charge accumulation is confined in the edges but the electric field is produced everywhere inside the macroscopic sample. As the relaxation of the electric field is infinite, this corresponds to the case without spin-flip scattering: $l_{sf} \rightarrow \infty$ of the previous section, but where $\partial \mu_0 /\partial y \ne 0$. In other words, the effect of the spin-Hall effect is equivalent to that of a Ferromagneti/Normal junction in a GMR experiment (especially in the lateral, or non-local, configuration): it puts out-of-equilibrium the two spin-populations up and down at the interface, as if an external spin pumping force \cite{Entropy} would be applied at the boundaries. The consequence is to induce a gradient of $\Delta \mu(y) \propto e^{-y/l_{sf}}$ over the spin-diffusion length.
 
 Independently, we can look at the approximation of small $\lambda_D$, without assuming constant conductivity. Multiplying Eq.(\ref{Screeningsf2}) by $\lambda_D^2$ yields:

\begin{equation}
\lambda_D^2 \left ( \frac{\partial^2 \delta n_{\updownarrow}}{\partial y^2} - \frac{\dot R}{2 D}  \right )- \delta n_{\updownarrow} -  \frac{1}{n_{\updownarrow}}  \frac{\partial \delta n_{\updownarrow}}{\partial y}  \int_y \delta n_{\updownarrow} \, dy' +  \frac{\lambda_D^2 }{kT} \left (n_{\updownarrow} \frac{\partial^2  \mu_{0 \updownarrow}}{\partial y^2} + \frac{\partial  \mu_{0 \updownarrow}}{\partial y} \, \frac{\partial  \delta n_{\updownarrow}}{\partial y}   \right ) =  \pm \frac{\lambda_D^2}{l_{sf}^2} \frac{\Delta \mu}{kT} ,
\label{Screeningsf6}
\end{equation}

In a metal, the Debye length $\lambda_D$ is of the order of a nanometer while the spin-flip relaxation length $l_{sf}$ is few tens of nanometers. To second order in $\lambda_D$, Eq.(\ref{Screeningsf6}) reduces to:

 \begin{equation}
 - \delta n_{\updownarrow} -  \frac{1}{n_{\updownarrow}}  \frac{\partial \delta n_{\updownarrow}}{\partial y}  \int_y \delta n \, dy'  \approx  0
 \label{Screeningsf7}
\end{equation}

Equation Eq.(\ref{Screeningsf7}) which describes the SHE with spin-flip scattering is the same as Eq.(\ref{ScreeningS4}) which describe SHE without spin-flip scattering. The properties discussed in the last section remain valid for the two stationary states.

To conclude this section, we can state that despite the existence of the GMR-like spin-accumulation at the interface over a distance $l_{sf}$, there is no qualitative change introduced by the spin-flip relaxation in the bulk SHE, as long as the spin-diffusion length is much larger than the Debye length $l_{sf} \gg \lambda_D$. In particular, in this limit, {\it the two stationary states cannot be discriminated by the spin-accumulation properties}.  

\section {Transport equations}

In this Section, the transport equations Eq.(\ref{OhmS2}) in the presence of the chemical potential Eq.(\ref{ChemPotS}) are analyzed. We show in Subsection {\bf A} that the transport equation is equivalent to the Dyakonov-Perel equations in the case of an electric field $E_{y \updownarrow} =0$, so that $\vec E_{\updownarrow} = \vec E$ is spin-independent. In Subsection {\bf B}, the transport equation Eq.(\ref{OhmS2}) are analyzed assuming that the two sub-systems can be treated separately. A generalized Dyakonov-Perel equation is obtained.

\subsection{Case of spin-independent electric field $\vec E$} 

If the electric potential $V(H_{so \updownarrow})$ is spin-independent, the chemical potential reads $\mu_{\updownarrow} = kT ln(n_{\updownarrow} ) + V + \mu_{0 \updownarrow}$. Inserted in Eq.(\ref{OhmS2}) we obtain, for $\vec \nabla \mu_{0 \updownarrow} = 0$:
\begin{equation}
\vec J_{\updownarrow}= q \eta n_{\updownarrow} \vec E - D \vec \nabla n_{\updownarrow} \pm \vec p \times  \left ( - q n_{\updownarrow} \eta_{so} \vec E +  D_{so} \vec \nabla n_{\updownarrow})\right )
\label{Ohm_mobil2}
\end{equation}
where $D_{so} = \eta_{so} kT$ is the spin-orbit diffusion constant.

Defining the asymmetry of charge carriers density between the two spin channels $\Delta n = n_{\uparrow} - n_{\downarrow}$ and $n = n_{\uparrow} + n_{\downarrow}$, we can define the charge current $\vec J_{c} = \vec J_{\uparrow} + \vec J_{\downarrow}$ and the spin current $\vec J_s = \vec J_{\uparrow} - \vec J_{\downarrow}$ by summing and subtracting the two Eqs.(\ref{Ohm_mobil2}):
 \begin{equation}
 \begin{array}{c}
\vec J_c= q \eta n \vec{ E} - D \vec \nabla  n  +  \vec p \times \left (- q \eta_{so} \Delta n \vec E + D_{so} \, \vec \nabla (\Delta n) \right) 
\\
\vec J_s=  q \eta \Delta n  \vec E - D \vec \nabla (\Delta n)  + \vec p \times \left ( -q \eta_{so} n  \vec E + D_{so} \, \vec \nabla n  \right)
\label{Ohm_SO}
\end{array}
\end{equation}
We can check that Eq.(\ref{Ohm_SO}) is equivalent to the Dyakonov-Perel equations written in the form proposed in reference \cite{Dyakonov,Dyakonov2}):
 \begin{equation}
 \begin{array}{c}
\vec J_c = \tilde \mu n \vec E + \tilde D \vec \nabla n + b \vec E \times \vec P + \delta \, \vec{rot}P \\
q_{ij} = - \tilde \mu E_i P_j - \tilde D \frac{\partial P_j}{\partial x_i} + \epsilon_{ijk} \left (b n E_k + \delta \frac{\partial n}{\partial x_k} \right).
\end{array}
\label{DP_True}
\end{equation}

Since we have $\vec{rot} (\Delta n \vec p) = - \vec p \times \vec \nabla (\Delta n)$, Eqs.(\ref{DP_True}) are equivalent to Eqs.(\ref{Ohm_SO}) of the two channel model providing the phenomenological constants of the model are defined as follows: the mobility of charge carriers is $\tilde \mu = q \eta$, the spin-orbit mobility is $b =- q \eta_{so}$,the diffusion constant is $\tilde D = - D$, the spin-orbit diffusion constant is $\delta = D_{so}$, and the spin-polarization is $\vec P = -\Delta n \vec p$.

The main solution of Eq.(\ref{Ohm_SO}) can be calculated in the following simple configuration in the bulk.
 Since the spin-Hall effect is characterized by zero charge accumulation at the edges $\delta n = 0$, the electric field along the axis $y$ also vanishes, $E_y = 0$, in the absence of other contributions. For the bulk approximation the diffusion terms and the spin accumulation $\Delta n$ vanish, and the DP equations reduces to $J_{x c} = q \eta n \, E_{x}^0$  and $J_{s y} =  q \eta_{so}  n \, E_{x}^0$ (and $J_{c x} = J_{s x} = 0$). {\it These equations necessarily lead to a non-vanishing pure spin-current} $J_{y \uparrow} = - J_{y \downarrow} \ne 0$.  
The spin-Hall angle is defined by the ratio of the spin currents $J_{sy}$ to the injected current $J_{c x}$: $\theta_{SH} = J_{sy}/J_{cx} = \eta_{so}/\eta$.  The form Eq.(\ref{Ohm_SO}) of the DP equations is probably responsible for the fact that the second stationary sate with $J_{\updownarrow sy} = 0$ (described below) has been overlooked. 

\subsection{Case of spin-dependent electric field $\vec E_{\updownarrow}$} 

We discuss now the case developed in this work, in which the electric field $E_{y \updownarrow}$ is formally defined in each spin-channel by to Gauss's law $\partial E_{y \updownarrow}/ \partial y = q\delta n_{\updownarrow}/\epsilon$. Introducing the chemical potential Eq.(\ref{ChemPotS})
into the transport equations Eq.(\ref{OhmS2}) leads to the following generalized DP equations:
\begin{equation}
\vec J_{\updownarrow}= q \eta n_{\updownarrow} \vec E_{\updownarrow} - D \vec \nabla n_{\updownarrow} \pm \vec p \times  \left ( - q n_{\updownarrow} \eta_{so} \vec E_{\updownarrow} +  D_{so} \vec \nabla n_{\updownarrow})\right )
\label{Ohm_mobil3}
\end{equation}
or:
 \begin{equation}
 \begin{array}{c}
\vec J_c= q \eta \left ( n_{\uparrow} \vec E_{\uparrow} + n_{\downarrow} \vec E_{\downarrow} \right ) - D \vec \nabla  n  +  \vec p \times \left (- q \eta_{so} \left ( n_{\uparrow} \vec E_{\uparrow} - n_{\downarrow} \vec E_{\downarrow} \right )  + D_{so} \, \vec \nabla (\Delta n) \right) 
\\
\vec J_s=  q \eta \left ( n_{\uparrow} \vec E_{\uparrow} - n_{\downarrow} \vec E_{\downarrow} \right )  - D \vec \nabla (\Delta n)  + \vec p \times \left ( -q \eta_{so}\left ( n_{\uparrow} \vec E_{\uparrow} + n_{\downarrow} \vec E_{\downarrow} \right )  + D_{so} \, \vec \nabla n  \right)
\label{DP_Gene}
\end{array}
\end{equation}
The particular case where $\vec E_{\updownarrow} = \vec E$ leads to the DP equations obtained in the previous subsection $\bf A$. 

Let us look at the simple situation considered in the previous subsection $\bf A$. Since the spin-Hall effect is characterized by the symmetry $\delta n_{\uparrow} = - \delta n_{\downarrow}$ (when $\eta$ is spin-independent), the relation $E_{y \uparrow} = - E_{y \downarrow}$ is verified when other contributions are absent. We then have $\Delta E = E_{y \uparrow} - E_{y \downarrow} \ne 0$ and $E_y = E_{y \uparrow} + E_{y \downarrow} =0$. 
For the bulk approximation the diffusion terms and the spin-accumulation $\Delta n$ vanish, such that the generalized DP equations Eqs.(\ref{DP_Gene}) are reduced to
$J_{c x} =  q \eta n \, E_{x}^0 + q \eta_{so} n E_{y \uparrow}$  and $J_{s y} = q \eta n E_{y \uparrow} - q \eta_{so}  n \, E_{x}^0$ ($J_{c y} = J_{s x} = 0$).

The stationarity condition of minimum power dissipation $J_{s y} = J_{s x} = 0$ {\it leads to the re-definition of the spin-dependent electric fields}: $E_{y \updownarrow} = \mp \theta_{SH} \, E_{x}^0 $. From the experimental point of view, the two results are very similar in both cases {\bf A} and {\bf B}, except that a {\it pure spin-voltage} $V_{y \updownarrow}$ is generated instead of a pure spin-current $J_{y \updownarrow}$, and the
{\it spin-Hall angle} $\theta_{SH} \equiv \eta_{so}/\eta$ is now measured as the ratio of the electric fields instead of the ratio of the electric currents. The transport equations Eq.(\ref{DP_Gene}) are more general as the usual DP equations since they can take into account both stationary states. In other terms, the systems described by Eqs.(\ref{DP_Gene}) are less constrained than the systems described by the DP equations.

\section{Conclusion}

We have analyzed the spin-Hall effect in the framework of the two spin-channel model depicted in Fig.1. It is shown that two stationary states can be defined. On the one hand, the stationary state is reached inside each spin channel and it is then defined by non-zero charge accumulation at the edges for both channels, and zero transverse current $J_{y \updownarrow} = 0$ in the bulk. On the other hand, the stationary state can be reached over the two undifferentiated spin channels, so that the stationary state is defined by zero charge accumulation, and non-zero pure spin-current $J_{y \updownarrow} = - J_{y \updownarrow}$ inside the bulk.  {\it  In both cases}, the total charge accumulation and total transverse electric field are zero, the spin-accumulation produced at the lateral edges is decreasing over the spin-diffusion length $l_{sf}$, and the spin-Hall angle is defined by the ratio $\theta_{SH}= \eta_{so}/\eta$. 
  
 However, it is shown that the transport equations  that correspond to the second situation (with the generation of pure spin-current in the bulk) are the Dyakonov-Perel equations while the former situation (without zero transverse current in the bulk) corresponds to a generalization of the Dyakonov-Perel equations, in which the electric field should be spin-dependent $\vec E_{\updownarrow}$. 

From the experimental point of view, the presence or absence of the charge accumulation $\delta n_{\updownarrow}$ or spin-dependent electric filed $E_{y \updownarrow}$ in each spin-channel is probably impossible to measure in the stationary state (only the total charge-accumulation and total field is accessible). Both quantities $\delta n_{\updownarrow}$  and $E_{y \updownarrow}$ could be fictitious (like typically the drift current and the diffusion current in a $n-p$ junction at equilibrium), or could be defined only during the transient states (at the femtoseconds time scale). This question is however beyond the scope of this study, as it should be performed in the framework of a short-time approach (transient states) that takes explicitly into account the spin-orbit relaxation time and the spin-flip relaxation time.

However, as the stationary state with non-zero pure spin-current $J_{y \updownarrow} \ne 0$ dissipates more than the other one, discriminating the two stationary states is easy in principle (but not in practice \cite{Otani}). Indeed, the ratio of the resistances measured with or without the pure spin-current on the same device is given by the square of the spin-Hall angle $\theta_{SH}$, as discussed in reference \cite {EPL2017}.

\end{document}